\PassOptionsToPackage{table}{xcolor}
\documentclass[sigconf,nonacm]{acmart}

\AtBeginDocument{%
  }

\setcopyright{acmlicensed}
\copyrightyear{2018}
\acmYear{2018}
\acmDOI{XXXXXXX.XXXXXXX}

\acmConference[Conference acronym 'XX]{Make sure to enter the correct
  conference title from your rights confirmation email}{June 03--05,
  2018}{Woodstock, NY}

\acmISBN{978-1-4503-XXXX-X/2018/06}
\usepackage{enumitem}
\usepackage{booktabs}
\usepackage{multirow}
\usepackage{makecell}
\usepackage{graphicx}
\usepackage[normalem]{ulem}
\usepackage{xcolor}
\definecolor{headerblue}{RGB}{228,238,249}
\definecolor{oursgreen}{RGB}{226,239,218}
\definecolor{datasetgray}{RGB}{242,242,242}

\begin{document}

\title{Hierarchical Latent Reasoning for LLM-based Recommendation}

\author{%
Peiyu Hu\textsuperscript{1,2,*},
Siying Gu\textsuperscript{2,*},
Weihai Lu\textsuperscript{3,*},
Zhuodong Liu\textsuperscript{4},
Yuntian Tang\textsuperscript{2},
Jiahao Liang\textsuperscript{2},
Yiying Xie\textsuperscript{2},
Jiang Rong\textsuperscript{2},
Zhaokai Luo\textsuperscript{2,\textdagger},
Zhiyong Wang\textsuperscript{2},
Jia Wang\textsuperscript{1,\textdaggerdbl}
}

\affiliation{%
  \institution{%
      \textsuperscript{1}Xi'an Jiaotong-Liverpool University \quad
    \textsuperscript{2}Xiaohongshu \quad
    \textsuperscript{3}Peking University \quad
    \textsuperscript{4}Beijing Jiaotong University
  }
  \country{}
}

\email{%
  peiyuhu30@gmail.com,
  sy.gu@stu.ecnu.edu.cn,
  weihai.lu@pku.edu.cn,
  zhuodong.liu@bjtu.edu.cn,
  jia.wang02@xjtlu.edu.cn,
}

\email{%
  {tangyuntian,liangjiahao1,yiyingxie, rongjiang,luozhaokai,sunzhenghuai}@xiaohongshu.com
}

\thanks{%
  \textsuperscript{*}These authors contributed equally to this work.\quad
  \textsuperscript{\textdagger}Team Leader.\quad
  \textsuperscript{\textdaggerdbl}Corresponding author: Jia Wang (jia.wang02@xjtlu.edu.cn).
}

\renewcommand{\shortauthors}{Hu et al.}

\begin{abstract}
Large Language Models (LLMs) have shown strong potential for
recommendation by leveraging their semantic understanding and
contextual modeling capabilities. Recent studies further introduce
reasoning mechanisms to improve user preference modeling. However,
explicit natural-language reasoning incurs substantial inference
overhead, whereas existing latent reasoning methods mainly focus on generating or
verifying intermediate states, leaving their layer-wise preference roles
and contributions insufficiently characterized. We propose
\textbf{HiLaR}, a \textbf{Hi}erarchical \textbf{La}tent
\textbf{R}easoning framework with layer-aware reinforcement optimization
for LLM-based recommendation. HiLaR
constructs temporal-guided hierarchical user preference
representations, aligns them with multiple LLM latent reasoning states,
and organizes the reasoning process from broad preferences to
fine-grained current intents. To further optimize the reasoning
trajectory, HiLaR combines final recommendation feedback with layer-aware process rewards derived from the marginal target-likelihood gain of each state. Experiments on four Amazon benchmark datasets show that HiLaR generally outperforms strong sequential, generative, and LLM-based
recommendation baselines. Ablation and sensitivity analyses further
verify the contribution of hierarchical representation learning, latent
alignment, and process-level optimization. Our code is available in \url{https://github.com/hupeiyu21/HiLaR}.
\end{abstract}




\maketitle

\section{Introduction}

Recommendation systems have become essential components of modern online services by connecting users with massive amounts of content~\cite{survey1,latentR3, gencdr, sid-survey, lu2026dealt}. Traditional recommendation systems typically follow a discriminative modeling paradigm, where user-item matching functions are learned to score and rank candidate items~\cite{GRU4Rec,kang2018self, liu2025coherency, li2026decoding}. However, such approaches mainly rely on user-item interaction signals, making it difficult to fully exploit rich item semantics and understand complex user preferences.

Large Language Models (LLMs)~\cite{llm1,qwen2} have recently advanced recommendation systems by enabling deeper semantic and preference modeling~\cite{zhang2024binllm,zhang2025collm}. Nevertheless, modeling how user interests evolve across multiple
granularities remains challenging. While recent methods introduce explicit natural-language reasoning (e.g., Chain-of-Thought) to achieve this, they incur substantial token overhead and redundancy during inference~\cite{llm-cot1,nye2021scratchpad}. This motivates latent reasoning approaches, which replace verbose text generation with continuous hidden states~\cite{hao2024coconut,deng2023implicit,zelikman2024quietstar, lu2026mm}.

\begin{figure}[t]
    \centering
    \includegraphics[
        width=0.93\columnwidth
    ]{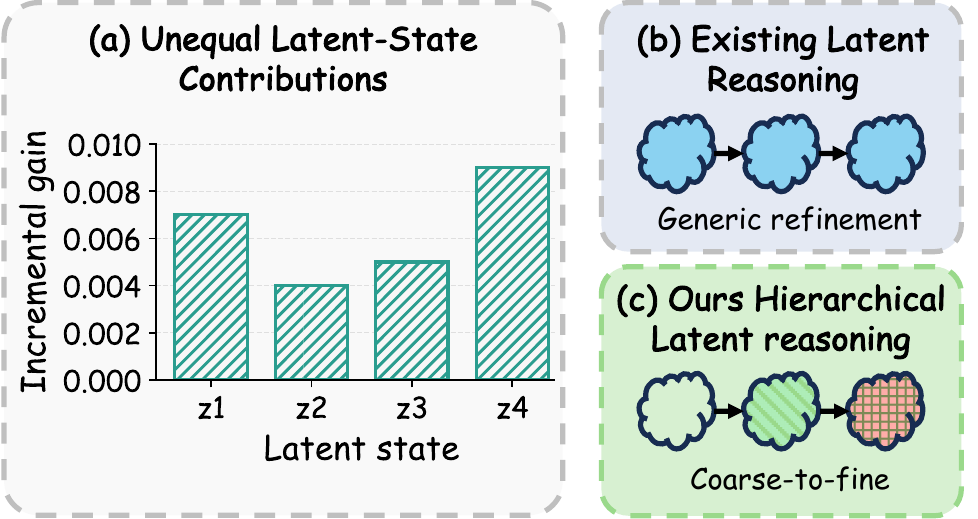}
    \caption{
    Motivation for HiLaR.
    (a) Different latent states exhibit varying marginal contributions to target-item generation; their differences are statistically significant under a repeated-measures ANOVA ($p<0.05$).
    (b) Existing methods do not explicitly ground their layer-wise roles in
    the multi-granularity structure of user preferences.
    (c) HiLaR introduces a temporally grounded, coarse-to-fine organization
    of latent reasoning states.
    }
    \label{fig:motivation}
\end{figure}

Pioneering works like LatentR$^3$~\cite{latentR3} have demonstrated the efficiency of sequential latent reasoning, typically treating intermediate states as general refinement steps (Fig. ~\ref{fig:motivation}(b)). However, our analysis reveals that individual states yield varying marginal contributions to target-item generation (Figure~\ref{fig:motivation}(a)). Conventional outcome-based rewards provide only shared trajectory-level feedback, failing to isolate step-wise contributions. Although VRec~\cite{vrec} mitigates this via intermediate verification, it assesses general reliability rather than assigning distinct temporal preference granularities or evaluating marginal gains. Because user histories contain preference signals ranging from stable
long-term interests to specific recent intents, we leverage this temporal structure as an inductive bias to organize the latent states into a coarse-to-fine hierarchy (Figure~\ref{fig:motivation}(c)). This systematic transition from broad preferences to fine-grained intents allows us to explicitly optimize each layer's marginal contribution through layer-aware process rewards.

Translating this conceptual hierarchy into a functional latent reasoning mechanism, however, presents two fundamental challenges.
(1) \textbf{task-grounded hierarchical modeling}: User histories
contain preference signals ranging from stable interests to recent
intents~\cite{kang2018self,chi2022long}, but latent states lack explicit
supervision that associates them with such coarse-to-fine preference
granularities.
(2) \textbf{layer-sensitive credit assignment}: Final rewards evaluate
the complete trajectory, while verification mainly assesses the
reliability of intermediate states. Neither directly quantifies the
marginal contribution of each state to target-item generation, making
multi-step latent optimization challenging under sparse feedback
~\cite{lightman2023verify,wang2024mathshepherd}.

To address these challenges, we propose \textbf{HiLaR}, a
\textbf{Hi}erarchical \textbf{La}tent \textbf{R}easoning framework with
layer-aware reinforcement optimization for LLM-based recommendation.
HiLaR consists of three components. First, Temporal-Guided Hierarchical
Quantization encodes time-aware user histories and applies multi-level
residual quantization, with progressively focused temporal windows
supervising the cumulative representation at each level. Second,
Hierarchical Latent Alignment Fine-tuning jointly optimizes target-item
generation and layer-wise alignment between LLM latent states and the
quantized preference representations. Finally, Hierarchical
Reward-Guided GRPO samples multiple reasoning trajectories and combines
final recommendation rewards with layer-wise alignment and marginal
target-likelihood gains to compute group-relative advantages. Together,
these components organize and optimize latent reasoning from broad
preferences to fine-grained current intents.

\begin{itemize}
    \item We propose \textbf{HiLaR}, which explicitly organizes latent
    reasoning states into a hierarchical structure for multi-granularity
    user preference modeling.

    \item We introduce \textbf{Temporal-Guided Hierarchical Quantization} and latent alignment to learn complementary coarse-to-fine preference
    representations.

    \item We develop \textbf{Hierarchical Reward-Guided GRPO}, which uses
    layer-wise target-likelihood gains to provide fine-grained feedback for
    latent reasoning optimization.

    \item Extensive experiments on multiple real-world datasets show
    that HiLaR outperforms strong baselines on most datasets and evaluation metrics.
\end{itemize}

\section{Related Work}

\textbf{LLM-based Recommendation.}
LLMs have been explored for recommendation through unified language-based
frameworks, instruction tuning, hybrid collaborative modeling, and
generative ranking~\cite{geng2022p5,cui2022m6rec,zhang2024instructrec,bao2023tallrec,wei2024llmrec,liao2024llara,zhang2024binllm,zhang2025collm,ji2024genrec,yue2023llamarec}.
Recent works further incorporate rationales, reasoning graphs, and
structured reasoning to improve preference modeling
~\cite{gao2023chatrec,wang2023llmrg,wang2024rdrec,bismay2025reasoningrec,llm4rec-reasoning1,llm4rec-reasoning3,llm4rec-reasoning4}.
However, explicit reasoning traces are costly to decode and may contain
redundant or weakly grounded information, motivating more efficient
reasoning mechanisms.

\textbf{Latent Reasoning and Process Supervision.}
Latent reasoning replaces discrete intermediate thoughts with continuous
hidden representations, building on chain-of-thought and scratchpad
reasoning~\cite{nye2021scratchpad,llm-cot1,llm-cot,zhu2025surveylatentreasoning}
and subsequent developments in implicit, pause-token, internal, and
recurrent reasoning~\cite{deng2023implicit,zelikman2024quietstar,hao2024coconut}.
For recommendation, LatentR$^3$ and VRec introduce continuous preference
reasoning and latent-state verification, while later studies explore
latent refinement, parallel trajectories, disentangled factors, and
adaptive reasoning depth~\cite{latentR3,vrec,liu2025lares,tang2026parallel,gao2026factorized,chen2026lasar}.
Process-supervision studies demonstrate the value of step-level feedback
for credit assignment and reasoning reliability
~\cite{lightman2023verify,wang2024mathshepherd,zheng2025processbench,luo2024improve}.
Different from these works, HiLaR imposes a temporally grounded
coarse-to-fine hierarchy and performs layer-aware optimization for
recommendation reasoning.

\section{Problem Formulation}
\label{ssec:problem_formulation}

Let $\mathcal{H}_u=[i_1,\ldots,i_T]$ denote the chronological
interaction history of user $u$, and let $i^+\in\mathcal{I}$ be the
next target item with canonical title $y^+$. Given $\mathcal{H}_u$, HiLaR
generates $K$ continuous latent reasoning states
$\boldsymbol{\tau}_u=(\tau_u^1,\ldots,\tau_u^K)$ and predicts the target
title:
\begin{equation}
p_{\theta}(y^+\mid\mathcal{H}_u)
=
p_{\theta}
\left(
y^+\mid\mathcal{H}_u,\boldsymbol{\tau}_u
\right).
\end{equation}
At inference time, catalog-constrained decoding produces an ordered
Top-$N$ recommendation list. The auxiliary quantization, alignment, and
reinforcement-learning objectives introduced below are designed to
improve target-item generation and organize the latent reasoning states
from coarse preferences to fine-grained intents.

\section{Method}

\begin{figure*}[ht]
  \centering
  \includegraphics[width=0.98\textwidth]{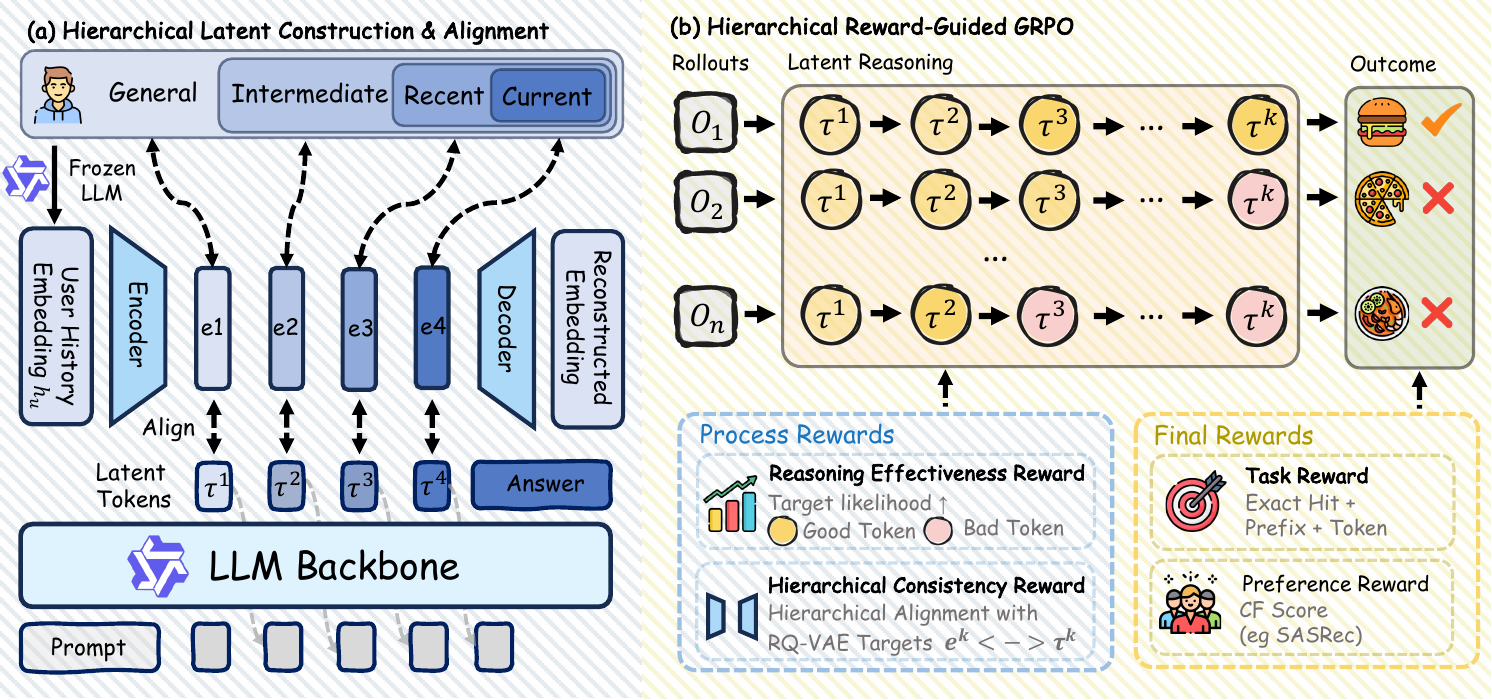}
  \caption{
    Overview of HiLaR. 
    (a) \textit{Hierarchical Latent Construction and Alignment} encodes
    time-aware user histories into hierarchical preference representations
    and aligns them with the latent reasoning states of the LLM.
    (b) \textit{Hierarchical Reward-Guided GRPO} samples multiple latent
    reasoning trajectories and optimizes them with process-level and final
    recommendation rewards.
    }
\label{fig:architecture}
\end{figure*}

To address the lack of structured preference modeling and layer-aware
optimization in existing latent reasoning recommenders, we propose
\textbf{HiLaR} (\textbf{Hi}erarchical \textbf{La}tent
\textbf{R}easoning), a hierarchical latent reasoning framework for
LLM-based recommendation. As shown in
Figure~\ref{fig:architecture}, HiLaR consists of three stages:
(1) \textit{Temporal-Guided Hierarchical Quantization}, which constructs
coarse-to-fine user preference representations from time-aware histories;
(2) \textit{Hierarchical Latent Alignment Fine-tuning}, which transfers
these representations to the LLM's latent reasoning states; and
(3) \textit{Hierarchical Reward-Guided GRPO}, which optimizes sampled
reasoning trajectories using final recommendation rewards and
layer-aware process rewards.

\subsection{Temporal-Guided Hierarchical Quantization}
\label{ssec:hierarchical_quantization}

User histories mix stable interests with recent intents~\cite{kang2018self,chi2022long}. To organize these preferences from coarse to fine, we introduce Temporal-Guided Hierarchical Quantization, which provides structured supervision for the latent reasoning states.

\textbf{User Representation Encoding.}
To provide hierarchical supervision for latent reasoning, the user
representation should preserve both item semantics and the temporal order
of historical behaviors. We therefore organize item information and
interaction timestamps into a time-aware prompt
$\mathcal{P}(\mathcal{H}_u)$ and use a frozen LLM to obtain the hidden
state at the end of the sequence:
\begin{equation}
\mathbf{h}_u =
\left[
\operatorname{LLM}_{\mathrm{frozen}}
\bigl(\mathcal{P}(\mathcal{H}_u)\bigr)
\right]_{\langle\mathrm{EndOfHistory}\rangle}.
\end{equation}
The quantizer is trained over all training users with shared
parameters, while each user obtains an individual representation
$\mathbf{h}_u$.

\textbf{Hierarchical Residual Quantization.}
Inspired by residual vector quantization
~\cite{van2017neural,lee2022autoregressive}, we decompose
$\mathbf{h}_u$ into multiple preference granularities using
$K$ shared codebooks. For user $u$, the
quantizer progressively removes the already-explained component from
the residual representation:
\begin{align}
\mathbf{q}_{u,k}
&=
\mathbf{E}_k(c_{u,k}), \qquad
c_{u,k}
=
\arg\min_{c\in\mathcal{C}_k}
\left\|
\mathbf{r}_{u,k-1}-\mathbf{E}_k(c)
\right\|_2^2, \\
\mathbf{r}_{u,k}
&=
\mathbf{r}_{u,k-1}-\mathbf{q}_{u,k},
\qquad
\mathbf{r}_{u,0}=\mathbf{h}_u .
\end{align}
The representation available after level $k$ is the cumulative
embedding
\begin{equation}
\mathbf{e}_{u,k}
=
\sum_{j=1}^{k}\mathbf{q}_{u,j}.
\end{equation}
Thus, each user has a personalized sequence
$\{\mathbf{e}_{u,k}\}_{k=1}^{K}$, although the codebooks are shared
across users. Later levels are encouraged to encode residual
preference information that is not captured by earlier levels.

\textbf{Temporal Hierarchical Supervision.}
Residual quantization alone does not determine what each level should
represent. We therefore divide each user's history into $K$ ordered
temporal windows,
$\mathcal{W}_{u,1},\ldots,\mathcal{W}_{u,K}$, from earlier to more
recent interactions. The supervision target becomes progressively more
focused as the quantization level increases:
\begin{equation}
\mathcal{Y}_{u,k}
=
\begin{cases}
\displaystyle\bigcup_{j=k}^{K}\mathcal{W}_{u,j}, & k<K,\\[6pt]
\{i^+\}, & k=K.
\end{cases}
\end{equation}
Accordingly, $\mathbf{e}_{u,1}$ is trained to reconstruct the user's
overall historical preference, while later representations focus on
recent behaviors and the target item. The temporal reconstruction loss
is
\begin{equation}
\mathcal{L}_{\mathrm{temp}}
=
\sum_{k=1}^{K}
\lambda_k
\operatorname{BCE}
\left(
\operatorname{Dec}_k(\mathbf{e}_{u,k}),
\mathcal{Y}_{u,k}
\right).
\end{equation}

The overall quantization objective combines representation
reconstruction, temporal supervision, and vector-quantization
regularization:
\begin{equation}
\mathcal{L}_{\mathrm{RQ}}
=
\left\|
\mathbf{h}_u-\mathbf{e}_{u,K}
\right\|_2^2
+
\mathcal{L}_{\mathrm{temp}}
+
\mathcal{L}_{\mathrm{VQ}},
\end{equation}
where $\mathcal{L}_{\mathrm{VQ}}$ contains the codebook and commitment
losses used for residual quantization. After training, the quantizer is
frozen, and the resulting personalized hierarchical representations
$\{\mathbf{e}_{u,k}\}_{k=1}^{K}$ are used to supervise the latent
reasoning states in the subsequent fine-tuning stage.

\subsection{Hierarchical Latent Alignment Fine-tuning}
\label{ssec:latent_alignment}

The hierarchical representations learned by the temporal-guided
quantizer provide useful preference targets, but they are not directly
constrained during LLM reasoning. To transfer this hierarchical
structure into the reasoning process, we align each latent reasoning
state with its corresponding preference representation.

Given a user history $\mathcal{H}_u$, the latent reasoning module
generates $K$ continuous states:
\begin{equation}
\boldsymbol{\tau}_u
=
\{\tau_u^1,\tau_u^2,\ldots,\tau_u^K\}.
\end{equation}

The $k$-th state is aligned with the $k$-th quantized representation
$\mathbf{e}_{u,k}$. When the two representations have different
dimensions, a learnable projection $\operatorname{Proj}$ maps the
reasoning state to the quantizer space:
\begin{equation}
\mathcal{L}_{\mathrm{align}}
=
\frac{1}{K}
\sum_{k=1}^{K}
\left\|
\operatorname{Proj}(\tau_u^k)
-
\mathbf{e}_{u,k}
\right\|_2^2.
\end{equation}

This layer-wise constraint encourages different reasoning states to
preserve the coarse-to-fine preference structure learned from user
histories. The SFT objective jointly optimizes recommendation
generation and hierarchical alignment:
\begin{equation}
\mathcal{L}_{\mathrm{SFT}}
=
\mathcal{L}_{\mathrm{rec}}
+
\lambda_{\mathrm{align}}
\mathcal{L}_{\mathrm{align}},
\end{equation}
where $\mathcal{L}_{\mathrm{rec}}$ denotes the target-item generation
loss.

\subsection{Hierarchical Reward-Guided GRPO}
\label{ssec:hierarchical_grpo}

Although hierarchical alignment provides useful supervision for latent
reasoning states, supervised fine-tuning relies on teacher-forced
generation and does not directly optimize sampled reasoning
trajectories. Moreover, final recommendation rewards are sparse and
cannot provide credit for individual latent states. We therefore use
GRPO~\cite{guo2025deepseek} to optimize both the final recommendation and the latent reasoning
process.

\textbf{Latent Reasoning Rollouts.}
To apply GRPO to continuous latent reasoning, we treat each latent
state as a stochastic reasoning action. Given a user history
$\mathcal{H}_u$, the $k$-th latent state is sampled from a Gaussian
policy conditioned on the history and previous latent states:
\begin{equation}
\tau_g^k
\sim
\mathcal{N}
\left(
\mu_{\theta,k}
(\mathcal{H}_u,\tau_g^{<k}),
\sigma_k^2\mathbf{I}
\right),
\qquad
k=1,\ldots,K,
\end{equation}
where $\mu_{\theta,k}$ is produced by the latent reasoning module and
$\sigma_k$ controls the exploration at the $k$-th reasoning level.

For each user, the policy samples $G$ rollout trajectories:
\begin{equation*}
\xi_g
=
\left(
\boldsymbol{\tau}_g^{1:K},O_g
\right),
\qquad
\boldsymbol{\tau}_g^{1:K}
=
(\tau_g^1,\ldots,\tau_g^K),
\end{equation*}
where $O_g$ is the generated item title. The latent trajectory probability is computed as
\begin{equation*}
\log \pi_\theta
\left(
\boldsymbol{\tau}_g^{1:K}\mid\mathcal{H}_u
\right)
=
\sum_{k=1}^{K}
\log
\mathcal{N}
\left(
\tau_g^k;
\mu_{\theta,k},
\sigma_k^2\mathbf{I}
\right).
\end{equation*}

For brevity, conditioning on $\mathcal{H}_u$ is omitted when clear from
context. The complete rollout probability is
\begin{equation*}
\log\pi_\theta(\xi_g)
=
\log\pi_\theta(\boldsymbol{\tau}_g^{1:K})
+
\log p_\theta(O_g\mid\boldsymbol{\tau}_g^{1:K}).
\end{equation*}

\textbf{Final Recommendation Rewards.}
Conventional recommendation RL often relies on a binary hit reward,
which provides no feedback when the generated item is not exactly
correct. This sparsity is particularly problematic for LLM-based
recommendation, where a generated title may be semantically close to
the target despite minor lexical differences. We therefore combine
exact matching with title-level similarity:
\begin{equation*}
R_{\mathrm{task}}^g
=
\lambda_{\mathrm{hit}}\mathbb{I}[O_g=y^+]
+
\lambda_{\mathrm{prefix}}R_{\mathrm{prefix}}^g
+
\lambda_{\mathrm{F1}}R_{\mathrm{F1}}^g,
\end{equation*}
where $y^+$ is the target title, $R_{\mathrm{prefix}}^g$ is the
normalized common-prefix score, and $R_{\mathrm{F1}}^g$ is the
token-level F1 score. The coefficients
$\lambda_{\mathrm{hit}}$, $\lambda_{\mathrm{prefix}}$, and
$\lambda_{\mathrm{F1}}$ are tunable hyperparameters that balance exact
matching and soft title similarity; their search ranges and selected
values are reported in the supplementary material.

The task reward may still overlook collaborative user preferences.
Thus, after mapping the generated title $O_g$ to an item $\hat{i}^g$,
we use a frozen collaborative recommender to compute:
\begin{equation*}
R_{\mathrm{pref}}^g
=
\operatorname{clip}
\left(
\frac{
s_{\mathrm{CF}}(u,\hat{i}^g)-\mu_u
}{
\sigma_u+\epsilon
},
-3,3
\right),
\end{equation*}
where $s_{\mathrm{CF}}$ is the user--item preference score, while
$\mu_u$ and $\sigma_u$ are computed from the candidate scores for user
$u$. The collaborative recommender is trained only on the training
split and remains frozen during GRPO.

\textbf{Process Rewards.}
Final rewards evaluate only the sampled output and provide limited
feedback about the latent reasoning process. We therefore introduce
dense rewards that measure both the effectiveness of the complete
trajectory and the marginal contribution of each latent state.

First, we measure whether the complete trajectory increases the
likelihood of the target title:
\begin{equation*}
R_{\mathrm{eff}}^g
=
\frac{1}{|y^+|}
\sum_{m=1}^{|y^+|}
\log p_{\theta}
\left(
y_m^+
\mid
y_{<m}^+,\mathcal{H}_u,
\boldsymbol{\tau}_g^{1:K}
\right).
\end{equation*}
This reward provides a dense signal even when the sampled title does not
exactly match the target.

However, $R_{\mathrm{eff}}^g$ only evaluates the complete trajectory
and cannot identify the contribution of each latent state. We therefore
compute the target log-probability after each latent prefix:
\begin{equation*}
\ell_{g,k}
=
\frac{1}{|y^+|}
\sum_{m=1}^{|y^+|}
\log p_{\theta}
\left(
y_m^+
\mid
y_{<m}^+,\mathcal{H}_u,
\boldsymbol{\tau}_g^{1:k}
\right),
\end{equation*}
where $\ell_{g,0}$ is computed without latent states. The marginal gain
of the $k$-th state is
\begin{equation*}
\Delta_{g,k}
=
\ell_{g,k}-\ell_{g,k-1}.
\end{equation*}
We use the clipped layer-wise gains to construct the process reward:
\begin{equation*}
R_{\mathrm{gain}}^g
=
\sum_{k=1}^{K}
\rho_k
\operatorname{clip}
\left(
\Delta_{g,k},
-b,b
\right),
\end{equation*}
where $\rho_k$ controls the relative importance of each reasoning level.

To prevent the policy from improving target likelihood while drifting
away from the hierarchical preference structure learned during
quantization, we further use an alignment reward:
\begin{equation*}
R_{\mathrm{align}}^g
=
\frac{1}{K}
\sum_{k=1}^{K}
\cos
\left(
\operatorname{Proj}(\tau_g^k),
\mathbf{e}_{u,k}
\right).
\end{equation*}
The complete process reward is
\begin{equation*}
R_{\mathrm{proc}}^g
=
R_{\mathrm{align}}^g
+
\lambda_{\mathrm{gain}}R_{\mathrm{gain}}^g.
\end{equation*}

\textbf{GRPO Optimization.}
The total reward combines final recommendation quality with
process-level reasoning feedback:
\begin{equation}
R^g
=
\lambda_{\mathrm{task}}R_{\mathrm{task}}^g
+
\lambda_{\mathrm{pref}}R_{\mathrm{pref}}^g
+
\lambda_{\mathrm{eff}}R_{\mathrm{eff}}^g
+
\lambda_{\mathrm{proc}}R_{\mathrm{proc}}^g.
\end{equation}

For the $G$ rollouts sampled for the same user, the group-relative
advantage is computed as
\begin{equation}
A_g
=
\frac{
R^g-\operatorname{mean}_{j=1}^{G}R^j
}{
\operatorname{std}_{j=1}^{G}R^j+\epsilon
}.
\end{equation}

Let $\pi_{\mathrm{old}}$ denote the rollout policy and
$\pi_{\mathrm{ref}}$ the frozen SFT policy. We define the trajectory-level
policy ratio as
\begin{equation}
r_g(\theta)
=
\exp\left(
\log\pi_\theta(\xi_g\mid\mathcal{H}_u)
-
\log\pi_{\mathrm{old}}(\xi_g\mid\mathcal{H}_u)
\right).
\end{equation}

The KL regularization contains both latent and text policies:
\begin{equation*}
D_{\mathrm{KL}}
=
D_{\mathrm{KL}}^{\mathrm{latent}}
+
D_{\mathrm{KL}}^{\mathrm{text}},
\end{equation*}
where $D_{\mathrm{KL}}^{\mathrm{latent}}$ is computed between the
corresponding Gaussian latent policies, while
$D_{\mathrm{KL}}^{\mathrm{text}}$ is computed over the autoregressive
token distributions of $\pi_\theta$ and $\pi_{\mathrm{ref}}$. Let
$\bar{r}_g(\theta)
=
\operatorname{clip}
\bigl(
r_g(\theta),1-\epsilon_{\mathrm{clip}},
1+\epsilon_{\mathrm{clip}}
\bigr)$
denote the clipped policy ratio. The GRPO objective is
\begin{equation}
\mathcal{L}_{\mathrm{GRPO}}
=
-\frac{1}{G}
\sum_{g=1}^{G}
\min\left\{
r_g(\theta)A_g,\,
\bar{r}_g(\theta)A_g
\right\}
+
\beta_{\mathrm{KL}}D_{\mathrm{KL}}.
\end{equation}
This objective improves recommendation quality while encouraging
hierarchical latent consistency and providing a layer-sensitive proxy for
credit assignment.

\section{Experiments}

\begin{table*}[t]
\centering
\resizebox{\textwidth}{!}{
\begin{tabular}{c|cccc|cccc|cccc|cccc}
\toprule
Method
& \multicolumn{4}{c|}{Toys}
& \multicolumn{4}{c|}{CDs}
& \multicolumn{4}{c|}{Games}
& \multicolumn{4}{c}{Instruments}\\

&
H@5 & H@10 & N@5 & N@10
&
H@5 & H@10 & N@5 & N@10
&
H@5 & H@10 & N@5 & N@10
&
H@5 & H@10 & N@5 & N@10\\

\midrule
\multicolumn{17}{c}{\textbf{Traditional Sequential Recommendation}}\\
\midrule

GRU4Rec
&0.0417&0.0564&0.0305&0.0352
&0.0481&0.0669&0.0365&0.0425
&0.0322&0.0517&0.0207&0.0270
&0.0766&0.0960&0.0630&0.0692\\

SASRec
&0.0601&0.0762&0.0458&0.0515
&0.0841&0.1054&0.0622&0.0691
&0.0416&0.0633&0.0285&0.0354
&0.0793&0.0950&0.0708&0.0758\\

LARES
&0.0823&0.1064&0.0615&0.0698
&0.0861&0.1022&0.0688&0.0743
&0.0635&0.0917&0.0458&0.0549
&0.1030&0.1245&0.0885&0.0953\\

\midrule
\multicolumn{17}{c}{\textbf{Generative Recommendation}}\\
\midrule

TIGER
&0.0786&0.0984&0.0572&0.0635
&0.1058&0.1215&0.0847&0.0906
&0.0563&0.0804&0.0395&0.0472
&0.0916&0.1112&0.0789&0.0846\\

RPG
&0.0815&0.1012&0.0598&0.0661
&0.1096&0.1250&0.0883&0.0939
&0.0589&0.0832&0.0411&0.0490
&0.0950&0.1148&0.0819&0.0878\\

\midrule
\multicolumn{17}{c}{\textbf{LLM-based Recommendation}}\\
\midrule

Base
&0.0203&0.0359&0.0128&0.0178
&0.0195&0.0252&0.0148&0.0167
&0.0236&0.0311&0.0190&0.0214
&0.0296&0.0411&0.0154&0.0192\\

CoT
&0.0261&0.0496&0.0153&0.0229
&0.0302&0.0406&0.0213&0.0246
&0.0120&0.0194&0.0082&0.0105
&0.0261&0.0452&0.0135&0.0199\\

AlphaRec
&0.0579&0.0893&0.0347&0.0448
&0.0479&0.0774&0.0278&0.0373
&0.0558&0.0893&0.0397&0.0515
&0.0813&0.1051&0.0564&0.0640\\

BIGRec
&0.0701&0.0931&0.0508&0.0582
&0.0757&0.0929&0.0616&0.0672
&0.0461&0.0709&0.0334&0.0414
&0.0938&0.1158&0.0807&0.0879\\

$D^3$
&0.0830&0.1026&0.0610&0.0674
&0.1122&0.1272&0.0906&0.0955
&0.0608&0.0860&0.0423&0.0505
&0.0984&0.1167&0.0848&0.0907\\

\midrule
\multicolumn{17}{c}{\textbf{Latent Reasoning Recommendation}}\\
\midrule

LatentR3
&0.0869&0.1135&0.0645&0.0745
&0.1131&0.1336&0.0911&0.0968
&0.0708&0.0964&0.0510&0.0578
&0.1032&0.1241&0.0894&0.0981\\

VRec
&0.0887&0.1132&0.0661&0.0740
&0.1146&0.1331&\textbf{0.0924}&0.0987
&0.0704&0.0991&0.0505&0.0603
&0.1056&0.1234&0.0916&0.0970\\

FLR
&\underline{0.0903}&\underline{0.1162}&\underline{0.0683}&\underline{0.0771}
&\underline{0.1167}&\underline{0.1454}&0.0908&\underline{0.0996}
&\underline{0.0745}&\underline{0.1041}&\underline{0.0537}&\underline{0.0627}
&\underline{0.1104}&\underline{0.1302}&\underline{0.0952}&\underline{0.1015}\\

HiLaR
&\textbf{0.0948}&\textbf{0.1213}&\textbf{0.0712}&\textbf{0.0801}
&\textbf{0.1176}&\textbf{0.1484}&\underline{0.0912}&\textbf{0.1012}
&\textbf{0.0768}&\textbf{0.1075}&\textbf{0.0548}&\textbf{0.0650}
&\textbf{0.1129}&\textbf{0.1320}&\textbf{0.0976}&\textbf{0.1038}\\

\bottomrule
\end{tabular}
}

\caption{
Overall performance comparison on four datasets. H@K and N@K denote Hit Ratio and NDCG at cutoff
$K$. Best and second-best results are shown in bold and underlined,
respectively. Improvements over the strongest baseline are evaluated
using paired $t$-tests at $p<0.05$.
}
\label{tab:main_results}
\end{table*}

We conduct experiments to answer the following research questions:

\begin{itemize}
    \item \textbf{RQ1 (Overall Performance):}
    Does HiLaR improve recommendation performance over
    traditional sequential, generative, and LLM-based recommendation
    baselines?

   \item \textbf{RQ2 (In-depth Analysis):}
    How do the main components of HiLaR contribute to its performance, and how
    sensitive is the model to key hyperparameters?

    \item \textbf{RQ3 (RL Training Dynamics):}
    Does layer-aware reward guidance improve the optimization dynamics
    of latent reasoning during GRPO training?

    \item \textbf{RQ4 (Latent Reasoning Analysis):}
    Do different latent reasoning states make distinct contributions, and how
    does HiLaR perform for users with different history lengths?
    
    \item \textbf{RQ5 (Inference Efficiency):}
    How does HiLaR compare with explicit and latent reasoning methods in
    terms of inference latency, GPU memory, and generation cost?
\end{itemize}

\subsection{Experimental Settings}

\textbf{Dataset.} Following LatentR3~\cite{latentR3}, we evaluate HiLaR on four
domain-specific Amazon datasets~\cite{ni2019justifying}: Toys, CDs,
Games, and Instruments. We adopt the same preprocessing and
chronological data-splitting protocol as LatentR3 to ensure a fair
comparison. Dataset statistics are reported in Table~\ref{tab:dataset}.

\begin{table}[t]
\centering
\small
\setlength{\tabcolsep}{5pt}

\begin{tabular}{c|rrrr}
\toprule
Dataset & Train & Valid & Test & Items \\
\midrule
Toys        & 53898 & 6737 & 6738 & 6299 \\
CDs         & 49251 & 6156 & 6158 & 5841 \\
Games       & 75175 & 9397 & 9397 & 5308 \\
Instruments & 66500 & 8312 & 8313 & 5030 \\
\bottomrule
\end{tabular}

\caption{
Statistics of the four Amazon datasets used in our experiments.
Train, Valid, and Test denote the numbers of training, validation, and
test instances, respectively.
}
\label{tab:dataset}
\end{table}

\textbf{Evaluation Metrics.}
We evaluate top-$N$ recommendation using HR@$N$ and NDCG@$N$,
where $N\in\{5,10\}$. All methods are evaluated under the
same candidate catalog, and HR@$N$ and NDCG@$N$ measure hit occurrence
and rank-aware recommendation quality, respectively.

\textbf{Baselines.}
We compare HiLaR with sequential recommenders (GRU4Rec~\cite{GRU4Rec}, SASRec~\cite{kang2018self}, latent-reasoning-enhanced (LARES~\cite{liu2025lares}), generative recommenders (TIGER~\cite{rajput2023recommender}, RPG~\cite{rpg}), LLM-based recommenders (Base, CoT~\cite{tsai2024leveraging}, AlphaRec~\cite{sheng2025language}, BIGRec~\cite{bao2025bi}, and $D^3$~\cite{bao2024decoding}), LatentR3~\cite{latentR3}, VRec~\cite{vrec}, and FLR~\cite{gao2026factorized}). LLM-based methods use the Qwen2.5-1.5B backbone~\cite{qwen2}, whereas sequential methods perform full-catalog ranking. Further details are provided in the Appendix.

\textbf{Implementation Details.}
Experiments use PyTorch on NVIDIA H20 GPUs with CUDA~12.6. We use a
frozen Qwen2.5-1.5B backbone and optimize trainable modules with AdamW
at a learning rate of $5\times10^{-6}$ and batch size $4$. The
$K=4$-level quantizer uses codebooks of size $256$ and is trained on
chronological equal-count windows using multi-label BCE with negative
sampling and level-specific decoders. SASRec is trained separately on
training interactions as the collaborative reward model; both are then
frozen. During SFT and GRPO, only the latent reasoning module and
projection layer are updated. We set $\lambda_{\mathrm{align}}=0.10$
for SFT and $\lambda_{\mathrm{proc}}=0.20$ for GRPO. GRPO samples $G=6$
rollouts with KL coefficient $0.05$ and layer-wise exploration noise
$(0.12,0.08,0.05,0.025)$. At inference, deterministic latent means and
catalog-constrained beam search are used with beam size $10$ and maximum
length $64$. Valid outputs are exactly matched to canonical catalog
titles. Main results are averaged over five random seeds; further
details appear in the supplementary material.

\subsection{Overall Performance (RQ1)}

Table~\ref{tab:main_results} compares HiLaR with traditional sequential,
generative, and LLM-based recommendation methods under the
Qwen2.5-1.5B backbone. HiLaR achieves the best performance on the Toys,
Games, and Instruments datasets, and obtains higher H@5, H@10, and N@10
than LatentR3 and VRec on CDs. VRec performs slightly better on CDs
N@5, indicating that HiLaR does not improve every metric uniformly.

These results suggest that latent reasoning and verification can improve
recommendation over standard LLM-based methods, while explicitly
organizing latent states into a hierarchical preference structure leads
to further gains. The improvements across different datasets also
indicate that complementary coarse-to-fine preference representations
are more effective than treating all latent states as homogeneous
refinement steps.

\subsection{In-depth Analysis (RQ2)}
\label{ssec:in_depth_analysis}

\textbf{Ablation Study.} We evaluate the contribution of each HiLaR component on the CDs and Toys datasets using the Qwen2.5-1.5B backbone. The variants w/o Temporal
Quant., w/o Hier. Align., and w/o GRPO remove the temporal-guided
residual quantization module, the hierarchical alignment loss, and
reinforcement learning, respectively. We further remove each reward
component, including the effectiveness, hierarchical alignment, and
collaborative preference rewards.

\begin{figure}[t]
    \centering
    \includegraphics[
        width=0.99\columnwidth
    ]{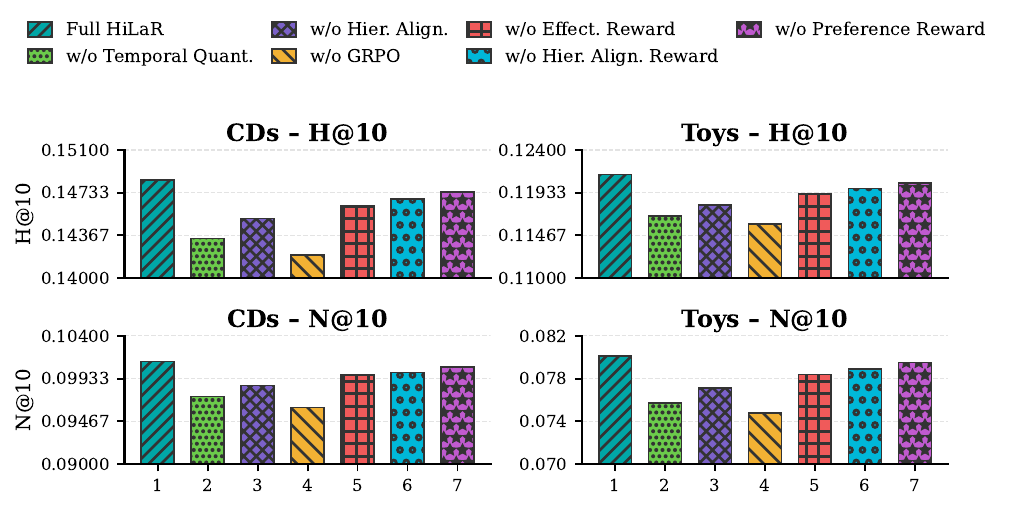}
    \caption{
    Ablation results on the CDs and Toys datasets using the
    Qwen2.5-1.5B backbone.
    }
    \label{fig:ablation}
\end{figure}

Figure~\ref{fig:ablation} reports H@10 and N@10 for the full model and
all ablated variants. Removing any component generally reduces
performance on both datasets. The largest drops occur after removing
Temporal Quantization or GRPO, highlighting the importance of hierarchical
preference construction and trajectory-level optimization. The reward
ablations further suggest that the three reward components provide
complementary guidance for latent reasoning optimization.

\begin{figure}[t]
    \centering
    \includegraphics[
        width=\columnwidth
    ]{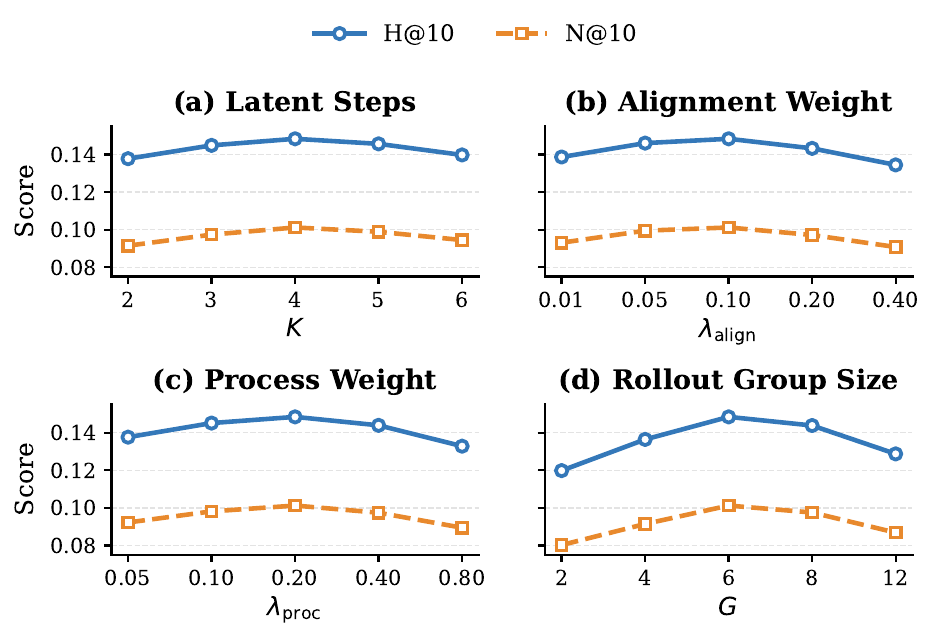}
    \caption{
    Hyperparameter sensitivity of HiLaR on the CDs dataset using the
    Qwen2.5-1.5B backbone.
    }
    \label{fig:hyperparameter}
\end{figure}

\textbf{Hyperparameter Sensitivity.}
Figure~\ref{fig:hyperparameter} examines the sensitivity of HiLaR to four
key hyperparameters on the CDs dataset using the Qwen2.5-1.5B backbone:
the number of latent reasoning steps $K$, the alignment weight
$\lambda_{\mathrm{align}}$, the process reward weight
$\lambda_{\mathrm{proc}}$, and the rollout group size $G$. We evaluate
H@10 and N@10 under different settings. HiLaR achieves its best or
near-best performance with $K=4$, $\lambda_{\mathrm{align}}=0.10$,
$\lambda_{\mathrm{proc}}=0.20$, and $G=6$. These results suggest that
moderate reasoning depth and balanced supervision and exploration provide
a favorable trade-off between hierarchical modeling and optimization
stability.

\subsection{RL Training Dynamics (RQ3)}

\begin{figure}[t]
    \centering
    \includegraphics[
        width=0.99\columnwidth
    ]{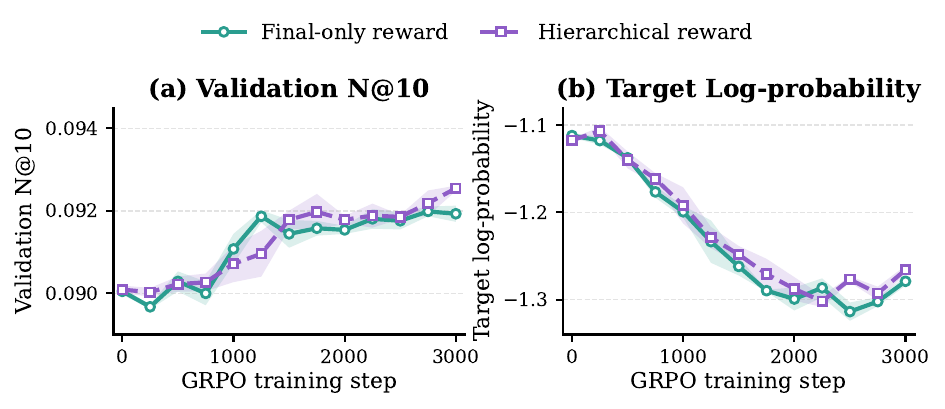}
    \caption{
    Training dynamics of GRPO with final-only and hierarchical rewards.
    The left panel reports validation N@10, while the right panel reports
    the target log-probability. Shaded regions indicate the standard deviation across five runs.
    }
    \label{fig:rl_training}
\end{figure}

Figure~\ref{fig:rl_training} compares the training dynamics of
final-only and hierarchical reward designs during GRPO optimization.
Both methods exhibit the characteristic fluctuations of reinforcement
learning, while the hierarchical reward gradually achieves better
validation N@10 in the later training stage. The target
log-probability also shows comparable but non-monotonic dynamics, with
the hierarchical reward obtaining a slightly better final value. These
results suggest that layer-aware process feedback can provide useful
optimization signals beyond the final recommendation reward, although
its effect is more apparent in the final ranking performance than in
the target likelihood alone.

\subsection{Latent Reasoning Analysis (RQ4)}
\label{ssec:case_study}

\begin{figure}[t]
    \centering
    \includegraphics[width=\columnwidth]{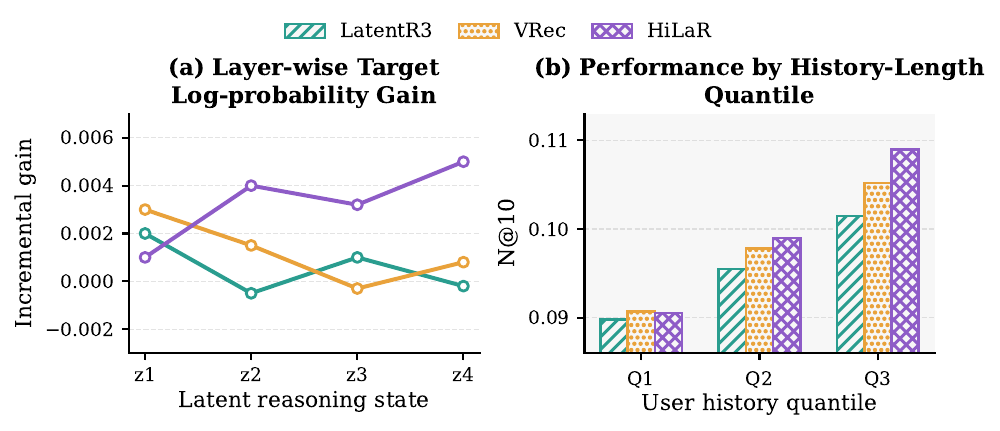}
    \caption{
    Case study of latent reasoning behavior. (a) Layer-wise target
    log-probability gain. (b) N@10 across user history-length quantiles,
    where Q1, Q2, and Q3 represent the shortest, middle, and longest
    history groups, respectively.
    }
    \label{fig:layer_user_analysis}
\end{figure}

Figure~\ref{fig:layer_user_analysis} analyzes the learned latent
reasoning behavior. As shown in Figure~\ref{fig:layer_user_analysis}(a),
HiLaR produces more differentiated layer-wise gains than LatentR3 and
VRec, suggesting that its latent states capture complementary
recommendation signals under the proposed coarse-to-fine supervision.

Figure~\ref{fig:layer_user_analysis}(b) reports N@10 across history-length
quantiles. While the methods perform similarly for short histories,
HiLaR shows a larger advantage for longer histories, indicating that
hierarchical latent reasoning is particularly useful when user behavior
contains richer and more diverse preference information.

\subsection{Inference Efficiency (RQ5)}

\begin{figure}[t]
    \centering
    \includegraphics[
        width=\columnwidth
    ]{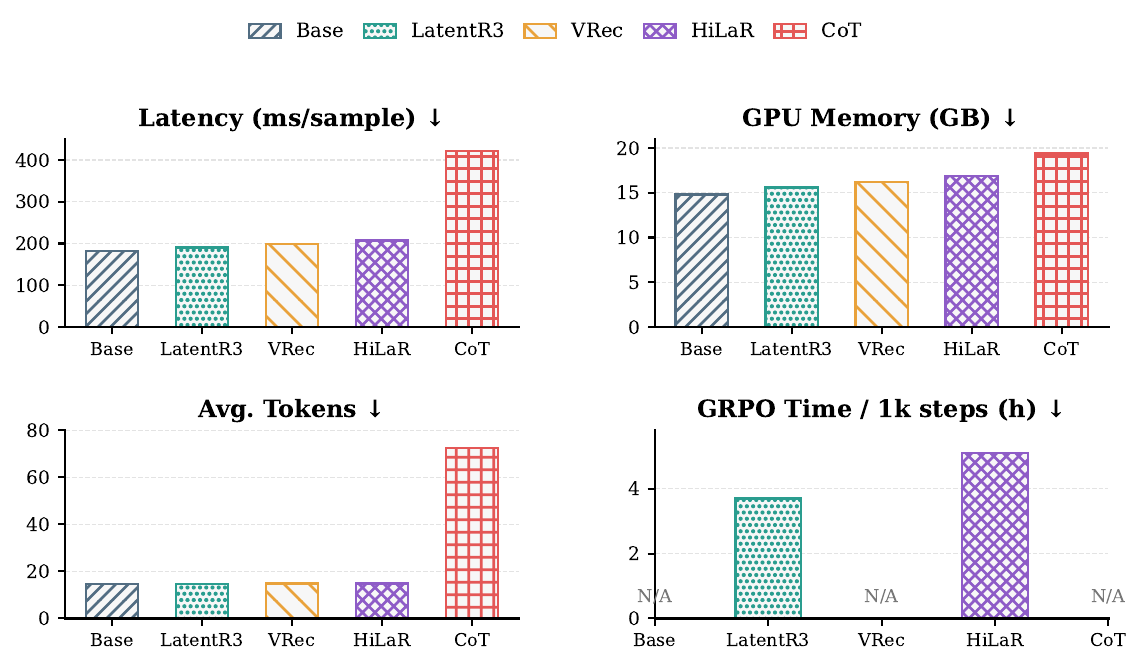}
    \caption{
    Efficiency comparison on CDs. Latency, GPU memory, and generated
    tokens are measured during inference. GRPO time is reported per
    1,000 update steps and is only applicable to GRPO-based methods.
    }
    \label{fig:efficiency}
\end{figure}

Figure~\ref{fig:efficiency} reports the absolute computational costs under
the same decoding configuration. LatentR3, VRec, and HiLaR introduce only
moderate inference overhead over Base, as their continuous reasoning
states require few additional generation steps. In contrast, CoT produces
more reasoning tokens, increasing latency and token consumption. HiLaR
incurs additional GRPO training time because its layer-wise rewards
evaluate multiple latent states, but this cost does not substantially
increase inference-time generation.

\section{Conclusion}

We proposed \textbf{HiLaR}, a hierarchical latent reasoning framework
for LLM-based recommendation. HiLaR constructs coarse-to-fine
preferences from time-aware histories, aligns them with LLM latent
states, and optimizes reasoning trajectories through layer-aware
reinforcement learning. Experiments on four datasets show that HiLaR
outperforms strong baselines on most metrics. Ablation and latent-state
analyses confirm the benefits of hierarchical representation, alignment,
and process-level optimization with moderate inference overhead.

\bibliographystyle{ACM-Reference-Format}
\bibliography{refs/refs}

\end{document}